\documentclass[12pt]{article} 
\usepackage{titlesec} 
\usepackage{placeins}
\usepackage{units} 
\usepackage[sharp]{easylist} 
\usepackage{array,booktabs}
\usepackage{colortbl}
\usepackage{abbrevs}
\usepackage{etoolbox}
\usepackage{booktabs}  
\usepackage{amsmath}
\usepackage{textcomp}  
\usepackage{lipsum}    
\usepackage{dcolumn}
\usepackage{bm}
\usepackage{cancel}
\usepackage{verbatim}
\usepackage{ifthen}
\usepackage{url}  
\usepackage{sectsty}
\usepackage{balance}  
\usepackage{graphicx} 
\usepackage{lastpage}
\usepackage[format=plain,justification=RaggedRight,singlelinecheck=false,
font=small,labelfont=bf,labelsep=period]{caption} 
\usepackage{fancyhdr}
\usepackage{color}
\usepackage{calc}
\usepackage{pdflscape}
\usepackage[utf8]{inputenc}
\usepackage[english]{babel}
\usepackage{natbib}
\usepackage{makecell}
\usepackage{comment}   
\usepackage{verbatim} 
\usepackage{lipsum, setspace} 
\usepackage{multirow}
\usepackage{pdfpages}
\usepackage{pdflscape}
\usepackage{pifont}
\usepackage{units}
\usepackage{mathtools}
\usepackage{dsfont}  
\usepackage{tcolorbox}
\usepackage[font=small,labelfont=bf]{caption} 
\usepackage{enumerate}
\usepackage{blindtext} 
\usepackage{tikz}
\usetikzlibrary{calc}
\usepackage[english]{babel} 
\usepackage{subcaption}   
\usepackage{caption}
\usepackage{setspace}
\usepackage{fullpage}

\newlength{\chaptercapitalheight}
\newlength{\chapterfootskip}
\newlength\graphht

\fancypagestyle{lscapedplain}{%
  \fancyhf{}
  \fancyfoot{
    \tikz[remember picture,overlay]
      \node[outer sep=1cm,above,rotate=90] at (current page.east) {\thepage};}

}

\newcommand{\bs}{ \mbox{\bf s}}

\newcommand{\bC}{ \mbox{\bf C}}

\newcommand{\bM}{ \mbox{\bf M}}

\newcommand{\btheta}{ \mbox{\boldmath $\theta$}}

\newcommand{\balpha}{ \mbox{\boldmath $\alpha$}}

\newcommand{\beq}{ \begin{equation}}
\newcommand{\eeq}{ \end{equation}}  
\newcommand{\beqn}{ \begin{eqnarray}}
\newcommand{\eeqn}{ \end{eqnarray}}

\newcommand{\bA}{\mathbf{A}}
\newcommand{\ba}{\mathbf{a}}

\newcommand{\bX}{\mathbf{X}}

\newcommand{\be}{\mathbf{e}}

\newcommand{\indep}{\stackrel{indep}{\sim}}
\newcommand{\iid}{\stackrel{iid}{\sim}}

\newcommand{\trips}[1]{{\left\vert\kern-0.25ex\left\vert\kern-0.25ex\left\vert #1
        \right\vert\kern-0.25ex\right\vert\kern-0.25ex\right\vert}}

\newcommand{\ind}{\mathrel{\text{\scalebox{1.07}{$\perp\mkern-10mu\perp$}}}}
\newtheorem{assumption}{Assumption}              

\newcolumntype{L}{@{}>{\kern\tabcolsep}l<{\kern\tabcolsep}}
\begin{document}

\vspace*{.5in}

\begin{center}
{\Large \textbf{Estimating intervention effects on infectious disease}}
\vspace{.2in}

{\Large \textbf{control: the effect of community mobility reduction on}}
\vspace{.2in}

{\Large \textbf{Coronavirus spread}}\\
\vspace{.4in}

{\large Andrew Giffin\footnote[1]{North Carolina State University}, 
Wenlong Gong$^1$, 
Suman Majumder\footnote[2]{Harvard T.H. Chan School of Public Health}, 
Ana G. Rappold\footnote[3]{Environmental Protection Agency},}
\vspace{.1in}

{\large Brian J.~Reich$^1$ and Shu Yang$^1$}\\

\vspace{.2in}
\today
\end{center}

\vspace{.1in}

\begin{abstract}\begin{singlespace}
\noindent Understanding the effects of interventions, such as restrictions on community and large group gatherings, is critical to controlling the spread of COVID-19. Susceptible-Infectious-Recovered (SIR) models are traditionally used to forecast the infection rates but do not provide insights into the causal effects of interventions. We propose a spatiotemporal model that estimates the causal effect of changes in community mobility (intervention) on infection rates. Using an approximation to the SIR model and incorporating spatiotemporal dependence, the proposed model estimates a direct and indirect (spillover) effect of intervention.  Under an interference and treatment ignorability assumption, this model is able to estimate causal intervention effects, and additionally allows for spatial interference between locations. Reductions in community mobility were measured by cell phone movement data. The results suggest that the reductions in mobility decrease Coronavirus cases 4 to 7 weeks after the intervention.
\vspace{12pt} \end{singlespace}\end{abstract}

\vspace{.1in}

\section{Introduction}

Since the Coronavirus exploded into a global pandemic, much research has been done to understand its epidemiological characteristics \citep[e.g.,][]{Li2020China, Livingston2020Italy} and quantify the effectiveness of various interventions to mitigate disease spread \citep[e.g.][]{Dandekar2020MLDL, Dehning2020Bayes, Cowling2020, Prem2020}. The traditional model for the progression of an infectious disease uses a set of differential  equations to decompose the population at risk into the number of susceptible (S), infected (I), or recovered (R) individuals and provide time-dependent trajectories of these compartments. Potential effects of interventions such as school closure, work from home, social distancing, etc. can be incorporated in the SIR model for the transitions between compartments. For example, \citet{Dandekar2020MLDL, Punn2020MLDL, Magdon-Ismail2020ML,lyu2020covid, kounchev2020tvbg} combine machine learning and/or deep learning with SIR model to make predictions of the spreading trend; while others, like \cite{Dehning2020Bayes, Mbuvha2020BayesSA, bradley2020joint}, adopt Bayesian approaches to provide uncertainty quantification of the prediction.

The SIR model \citep{kermack1927contribution} and the above variants do not account for the spatial distribution of the number of susceptible, infected, or recovered people. A number of attempts have been made to modify the SIR model to incorporate this spatial distribution. \cite{reluga2004two} and \cite{burger2009modelling} consider diffusion of the infected population by modeling the individual agents, while a number of works \citep[e.g.,][]{chinviriyasit2010numerical, hilker2007diffusive, robinson2012spatial, wang2010cross} consider Brownian motion-like movement for subgroups of people belonging to each compartment. Several works \citep{berres2011fully, milner2008sir, capasso1994asymptotic} use cross-diffusion, an assumption where the susceptible move away from an increasing gradient of the infected to bring in a spatial distribution to the SIR model. Models that assume that patches of population have different infection parameters and are connected by travel or migration have also been proposed \citep{arino2003multi,hyman2003modeling,sattenspiel1995structured,sattenspiel2003simulating,lee2012effect,lee2015role,lee2015spatial}.

Our method is based on the spatial SIR model of \cite{paeng2017continuous}. Among the several generalizations of the SIR model they propose is a model discrete in time and with space partitioned into distinct areas (e.g., counties).  The number of individuals in each compartment and area evolves over time using the same mechanics as the standard SIR model. A distance metric is defined between counties, allowing the number of infected in county $j$ to account for dependence across nearby areas. 

Our statistical contribution is using the SIR model as a stepping stone towards a spatiotemporal statistical model that permits causal inference without including differential-equation solvers. The result is a model with a very different flavor than the standard SIR model. In particular, while the goal of an SIR model is often to forecast the severity of a disease over time, the model presented here is designed to tease out causal effects of interventions. 

Integrating causal inference into this spatial framework presents some challenges. Existing SIR models are generally used for prediction and forecasting, but not for ascertaining causal effects of interventions. Causal methods generally require adjusting for confounders that are related to both intervention and response, which allows for identification of the causal intervention effect as opposed to the observed association between intervention and response. However, if interventions at one location affect the response at other locations -- a phenomenon known as interference -- then the standard causal framework which adjusts only for local confounding is rendered ineffective. With a virus that spreads across space, clearly some degree of interference is involved, and a causal analysis must take this into account. This is typically done by placing assumptions on the form of interference.  Commonly used assumptions on interference \citep[for a recent review, see][]{reich2020review} include ``partial interference'' in which the population is divided into isolated clusters \citep{halloran1991study, sobel2006randomized}, as well as ``network interference'' in which interference is allowed along a pre-specified network, and ``spatial interference'' in which interference is tempered by the distance between locations \citep{giffin2020generalized}. This paper uses a parsimonious assumption on interference, which allows for interference from neighboring counties only. This can be thought of as a simple version of both the spatial and network interference assumptions. Intervention effects are then divided between direct (within county) interventions and indirect (between county) interventions. 

The framework that we propose approximates the standard SIR model, in a way that allows for estimation of the causal effects of interventions. In particular, we use the fact that infection rates are likely smooth across space, to obtain an approximation to the rate of infection at across space and time. This approximation is a function of the intervention variables, covariates, and a spatiotemporal model, and allows us to fit the observed data and obtain intervention effect estimates.

The remainder of the paper proceeds as follows.  The motivating data are described in Section \ref{s:data}.  Section \ref{s:method} introduces the statistical method and corresponding theoretical properties. The method is evaluated using a simulation study in Section \ref{s:sim} where we show that the approximate model can recover causal effects of data generated from the SIR model. 
Section \ref{s:app} applies the proposed model to estimate the effect of decreases in mobility on Coronavirus cases, finding that decreases in mobility cause a decrease in local Coronavirus cases 4-7 week later.  Section \ref{s:discussion} concludes.

\section{Data description}\label{s:data}

We estimate the effect of mobility on the number of observed Coronavirus cases. There are three primary types of data for analysis: intervention (mobility reduction), response (Coronavirus cases) and covariates. All variables are defined temporally by the week ($t$) and spatially by the county ($j$).  Data from March 6, 2020 through October 8, 2020 for 3,137 counties or county equivalents are used in this study.

The response data $Y_j(t)$ are new, recorded cases of Coronavirus in a given county/week. These data are taken from the publicly available Johns Hopkins University Coronavirus Resource Center \citep{dong2020interactive}, which aggregates Coronavirus case counts and provides a daily, cumulative count of cases in each county. 

The intervention data are publicly available measures of mobility -- as measured by Google devices and provided by \cite{googleMobilityData} -- which have been shown to have strong associations with Coronavirus case data \citep{chang2020mobility,yilmazkuday2020stay}.  We use an aggregate mobility measure that includes the categories: ``retail and recreation'', ``grocery and pharmacy'', ``transit station'', ``workplace'',  and ``residential'' mobility. These measure both the number of visits to such locations as well as the length of stay in comparison to baseline. These data are given as percentage change from a baseline level which was taken over January 3, 2020-February 6, 2020. Intuitively, in the months since the pandemic began, ``retail and recreation'', ``grocery and pharmacy'', ``transit station'' and ``workplace'' mobility have decreased substantially from the baseline and ``residential'' mobility (i.e., amount of time spent at home) has increased. The specific metric that we will use as intervention $A_j(t)$ for a given county/week is the mean of percentage change from baseline of ``retail and recreation'', ``grocery and pharmacy'', ``transit station'', ``workplace'', and \textit{negative} ``residential'' mobility. 

Because of privacy concerns, county/days with too few users are not provided by Google. When a subset of the categories are provided for a given county/day, the mean of the available categories is taken. When none of the categories are provided for a given count/day, we impute the missing value from any available first and second degree neighbors (with values weighted by $1/$distance, where weights are summed to one). 

In addition to intervention and response we include a number of important covariates. From the 2016 American Community Survey we have poverty rate, population density, median household income, and total population, and from the U.S.~Census we have median age and the number of foreign born residents \citep{ACS}. The Bureau of Labor Statistics provides the number of people employed in healthcare and social services, classified as North American Industry Classification System, Sector 62) \citep{BLS}. The Environmental Systems Research Institute (ESRI) provides publicly available data on the number of hospital beds and the number of ICU beds \citep{ESRI}. Lastly, because \cite{Wu2020.04.05.20054502} shows that PM$_{2.5}$ (particulate matter smaller than 2.5 micrometers) may interact significantly with COVID-19, we include county level PM$_{2.5}$ estimates from 2016 \citep{alexeeff2015consequences,chudnovsky2012prediction}. We also include non-static county-level daily meteorological data, as this could inform levels of mobility and spread of disease. We use daily mean, maximum and minimum temperature (degrees Celsius), relative humidity and mean dew point (degrees Celsius) from NOAA's Global Surface Summary of the Day using the ``GSODR'' R package \citep{sparks2017gsodr}. Then we obtain county-level data by predicting at the mean county latitude and longitude coordinates from the 2010 Census using thin plate spline regression from the ``fields'' R package \citep{nychka2014fields}.

\section{Main methodology}\label{s:method}

\subsection{Notation}

Let $Y_{j}(t)$ be the number of new cases of Coronavirus reported during week $t\in\{1,...,T\}$ and county $j\in\{1,...,J\}$, where $T=31$ and $J =$ 3,137. 
For county $j$, denote $N_j$ as the population size and $\bX_j(t)$ as a vector of covariates for week $t$.  The direct intervention variable, $A_{j}(t)$, is the mobility percentage change from baseline for county $j$ and week $t$. In addition to the direct intervention, we allow for an indirect (spillover) intervention $\tilde{A}_j(t)$. This variable captures the interventions received by neighboring counties. Including this terms allows for interference between neighbors, since the response at county $j$ can now be impacted by the interventions in surrounding counties. The spatial configuration of the counties is summarized by their adjacency, with $c_{jj}=0$, $c_{jk}=1$ if counties $j$ and $k$ are adjacent and $c_{jk}=0$ otherwise. $\tilde{A}_j(t)$ is defined as the mean direct intervention over the $m_j=\sum_{k=1}^J c_{jk}$ adjacent regions: ${\tilde A}_{j}(t)=\sum_{k=1}^{J}c_{jk}A_j(t)/m_j$. Similarly, the mean of the adjacent covariate values is ${\tilde \bX}_{j}(t)=\sum_{k=1}^{J}c_{jk}\bX_j(t)/m_j$.

\subsection{Conceptual SIR model}\label{s:SIR}

To motivate the proposed statistical framework, we begin by describing the discrete-time, spatial SIR model proposed in \cite{paeng2017continuous}.  Let $S_{j}(t)$, $I_{j}(t)$ and $R_{j}(t)$ be the number of susceptible, infected and recovered individuals in county $j$ at time $t$. These three states evolve over time but always satisfy the constraint that $N_j=S_{j}(t)+I_{j}(t)+R_{j}(t)$. For each county $j$ the state evolution is similar to the classical SIR model \citep{kermack1927contribution}:
\begin{align}\label{e:SIR}
    \begin{split}
        S_j(t+1) - S_j(t) &= - \lambda_j(t),\\
        I_j(t+1) - I_j(t) &= \lambda_j(t) - \gamma I_j(t),\\
        R_j(t+1) - R_j(t) &= \gamma I_j(t),
    \end{split}
\end{align} 
where $\gamma_j>0$ is the recovery rate and $\lambda_j(t) = \beta\frac{S_j(t)}{N_j}\left\{\sum_{k=1}^JW_{jk}I_k(t)\right\}$ is the rate new infections in county $j$ and week $t$. Here $W_{jk}$ is proportional to the contact rate between individuals in region $j$ with individuals in region $k$.

We expand on this by allowing the infection rate $\beta_j(t)>0$ to vary with region and time:
\begin{equation}\label{e:lambda}
 \lambda_j(t) = \beta_j(t)\frac{S_j(t)}{N_j}\left\{\sum_{k=1}^JW_{jk}I_k(t)\right\}.
\end{equation}
This allows us to model spatiotemporal variation in $\beta_j(t)$ as a function of the covariates $\bX_j(t)$ and direct and indirect intervention variables $A_j(t)$ and $\tilde{A}_j(t)$. In the absence of other information on connectivity, we simply set $W_{jj} = (1-\phi)$ and $W_{jk}=\phi\cdot c_{jk}/m_j$ for $\phi\in[0,1]$ so that large $\phi$ leads to strong spatial dependence and vice versa. 

A major difficulty in using model (\ref{e:SIR}) is that we do not observe the latent states $S_j(t)$, $I_j(t)$ or $R_j(t)$.  We learn about these latent processes via the reported number of new cases $Y_j(t)$. Further complicating the statistical model is that there may be under-reporting and a lag time between an increase in the true and reported infection rate due to the latency of the disease.  Because of these issues, we link $Y_j(t)$ and $\lambda_j(t)$ in the SIR model by the over-dispersed Poisson distribution
\begin{equation}\label{e:EY}
 Y_{j}(t)|\lambda_j(t),g_j(t) \indep \mbox{Poisson}\left[ p \exp \{ g_j(t) \}\lambda_j(t-l)\right],
\end{equation}
where $p\in(0,1)$ accounts for under-reporting, $g_j(t)\iid\mbox{Normal}(0,\tau^2)$ captures over-dispersion, and $l\in\{0,1,2,...\}$ is the reporting lag.

\subsection{Gaussian approximation}\label{s:approx}

The SIR model in Section \ref{s:SIR} is an elegant way to model and forecast the spread of a disease through a population. However, the solution to the difference equations is complex, which hinders the ability to estimate model parameters in a statistical manner. Previously methods have been proposed to mimic the mechanistic model to provide realistic forecasts, e.g., \cite{buckingham2018gaussian}. Here, we propose an approximation to the SIR model to allow for estimation of intervention effects on local infection rates. 

The key insight is that the rate of new infections in (\ref{e:lambda}) can be re-expressed as 
\begin{eqnarray}\label{e:lambda2}
  \lambda_j(t) &=& \beta_j(t)\exp\{\theta_j(t)\} + v_j(t), \label{e:lambda2parts}\\
  \exp\{\theta_j(t)\} &=& S_j(t)I_j(t)/N_j, \label{e:lambda2partsb} \\ 
  v_j(t) &=& \beta_j(t)\phi\frac{S_j(t)}{N_j}\sum_{k=1}^J\frac{c_{jk}}{m_j}\{I_k(t)-I_j(t)\}. 
  \label{e:lambda2partsc}
\end{eqnarray}
Both $\theta_j(t)$ and $v_j(t)$ in (\ref{e:lambda2parts}) are latent spatial and temporal processes due to the unobserved $S_j(t)$ and $I_j(t)$, which we approximate by spatiotemporal models.

We model $\theta_j(t)$ as a separable spatiotemporal continuous autoregressive model \citep{stein2005space}.   Denote $\btheta(t) = \{\theta_1(t),...,\theta_J(t)\}$ and $\btheta = \{\btheta(1)^\top,...,\btheta(T)^\top\}^\top$.  Then the spatiotemporal conditional autoregressive model (STCAR) for $\btheta$ is multivariate normal with mean zero and covariance $\sigma^2\Sigma(\rho_t)\otimes \Omega(\rho_s)$.  Spatial dependence is governed by $\Sigma(\rho_s) = (\bM-\rho_s\bC)^{-1}$, where $\bM$ is diagonal with diagonal elements $m_j$ and the $(j,k)$ element of $\bC$ is 1 if sites $j$ and $k$ are adjacent and zero otherwise. If $\btheta(t)\sim\mbox{Normal}\{0,\sigma^2\Sigma(\rho_s)\}$, then $\theta_j(t)|\theta_k(t) \mbox{\ for all\ \ } k\ne j \sim\mbox{Normal}(\rho_s\sum_{k\sim j}\theta_k(t)/m_j,\sigma^2/m_j)$, so that $\rho_s$ determines the strength of spatial dependence and $\sigma$ is the scale parameter.  We denote this model as $\btheta(t)\sim\mbox{CAR}(\sigma,\rho_s)$.  Similarly, temporal dependence is governed by $\Omega(\rho_t)$, which has the same form as $\Sigma(\rho_s)$ except for temporal adjacency structure with time $t$ having $t-1$ and $t+1$ considered neighbors.  We refer to this as the $\mbox{STCAR}(\sigma,\rho_s,\rho_t)$ model.

The second term in (\ref{e:lambda2}), $v_t(t)$, sums over the $m_j$ regions adjacent to region $j$, and is a function of the local differences, $I_k(t)-I_j(t)$.  Assuming the number of infected individuals varies smoothly across space, these local differences should be small.  Therefore, one approximation we consider is simply setting $v_j(t)=0$ for all $j$ and $t$.  For cases where these terms cannot be removed, we note that because they are functions of local differences they should be less spatially correlated than the spatial process itself, and thus a second approximation is $\log \{ v_j(t) \} \iid\mbox{Normal}(\mu_v,\sigma_v^2)$.

A crucial point is that neither our model for $\theta_t(\bs)$ nor $v_t(\bs)$ attempt to retain the mechanistic properties of $I_j(t)$ and $S_j(t)$. Therefore, this model would not provide reliable forecasts about future disease spread.  However, forecasting is not our objective.  Our aim to is provide a computationally feasible method to use observed data to estimate effects of intervention variables on local infection rates.  The following section verifies that the approximation indeed has a causal interpretation under standard assumptions from causal inference, and proposes an estimation procedure for these effects.
 
For the final component, modeling the spatiotemporal infection rate $\beta_j(t)$, we regress $\log\{\beta_j(t)\}$ on covariates $\bX_j(t)$ and $\tilde{\bX}_j(t)$ as well as mobility variables $A_j(t)$ and $\tilde{A}_j(t)$ as
\begin{equation}\label{e:beta}
     \eta_j(t) = \log\{\beta_{j}(t)\} =  \alpha_0+ \bX_j(t)^\top\balpha_1+{\tilde \bX}_j(t)^\top\balpha_2 + A_j(t)\delta_1+{\tilde A}_j(t)\delta_2\nonumber\\
\end{equation}
where $\delta_1$ and $\delta_2$ quantify the direct and indirect (spillover) causal effects of mobility on infection rate.  The covariate vectors $\bX_j(t)$ and ${\tilde \bX}_j(t)$ can include both the original covariates or covariates derived as propensity scores (Section \ref{s:PO}).  

Since it is impossible to identify the reporting rate $p$ in (\ref{e:EY}) and intercept terms $\alpha_0$ and $\mu_v$ we fit the final model
\begin{eqnarray}
  Y_{j}(t) | g_j(t),\btheta,v_j(t) &
     \indep &
     \mbox{Poisson} \left[ \exp \{ g_j(t) + \eta_j (t-l)  + \theta_j(t-l) \} + 
     \exp\{{\tilde v}_j(t)\} 
     \right] \label{e:finalmodel}\\
    \eta_j(t) &=& \alpha_0+ \bX_j(t)^\top\balpha_1+{\tilde \bX}_j(t)^\top\balpha_2 +A_j(t)\delta_1+{\tilde A}_j(t)\delta_2\label{e:betaFormInFullSpec}\\     
g_j(t)&\iid&\mbox{Normal}(0,\tau^2)       \nonumber\\
  \btheta &\sim& \mbox{STCAR}(\sigma^2,\rho_s,\rho_t)\nonumber\\
  {\tilde v}_j(t) &\iid&\mbox{Normal}(\mu_v,\sigma_{\tilde v}^2).\nonumber
 \end{eqnarray}
The final term in (\ref{e:finalmodel}) combines the overdispersion term $g_j(t)$ and the nugget term $v_j(t)$.  When the dimensions of the covariates are large, we propose dimension reduction for $\log\{\beta_j(t)\}$ using the generalized propensity score \citep{imbens2000role} in Section \ref{s:PO}. To complete the Bayesian model, we specify noninformative independent Normal$(0,10^2)$ priors for $\alpha_0$, $\delta_1$, $\delta_2$, $\mu_v$ and the elements of $\balpha_1$ and $\balpha_2$, and noninformative priors for covariance parameters $\sigma^2,\tau^2,\sigma_{\tilde v}^2\sim\mbox{InvGamma}(0.1,0.1)$ and $\rho_s,\rho_t\sim\mbox{Uniform}(0,1)$.

\subsection{The potential outcomes framework}\label{s:PO}

In this section, we provide a causal interpretation of $\delta_1$ and $\delta_2$ under the potential outcomes framework \citep{rubin1974estimating}. To ease discussion, let $Y_j(t)$ be the total number of cases, rather than reported number (i.e., $p=1$). We use the potential outcomes framework to define the causal effect of mobility on infection rate. We use overbar to denote all history. Let $\bar{\ba}_j(t) = (a_j(1),\ldots,a_j(t))^\top$ be the trajectory of mobility level at region $j$ through time $t$. Let $\bar{\ba}(t) = (\bar{\ba}_1(t),\ldots,\bar{\ba}_J(t))$ be the trajectory of mobility level for all regions through time $t$.  Define $Y_j^{\bar{\ba}(t)}(t)$ to be the (possibly counterfactual) number of new cases in region $j$ at time $t$ had all the regions controlled the mobility level at $\bar{\ba}(t)$ through time $t$. 

Assume the potential outcome model for the SIR model with $\theta_j(t) = \log \{ S_j(t)I_j(t)/N_j \}$ and $v_j(t)=\beta_j(t)\frac{S_j(t)}{N_j}\sum_{k=1}^J\frac{c_{jk}}{m_j}\{I_k(t)-I_j(t)\}$ as in  (\ref{e:lambda2partsb}) and (\ref{e:lambda2partsc}), and 
\begin{align} \label{eq:M1}
Y_j^{\bar{\ba}(t)} &\sim \text{Poisson} \{\lambda_j^{\bar{\ba}(t)}\}, \nonumber \\
 \lambda_j^{\bar{\ba}(t)}(t) &=  \beta_j^{\bar{\ba}(t)}(t)   \exp \{ \theta_j(t) \} + v_j(t),  \\ 
\log\{\beta_j^{\bar{\ba}(t)}(t)\} &= 
a_j(t)\delta_1+{\tilde a}_j(t)\delta_2 + h \{ \bar{\bX}(t), \bar{\mathbf{S}}(t-1), \bar{\mathbf{I}}(t-1) \}.  \nonumber 
\end{align}
Consider two regimes $\bar{\ba}(t)$ and $\bar{\ba}'(t)$, where all components are the same except for $a_j(t) = a_j'(t)+1$, and $v_j(t) = 0$. Then under model (\ref{eq:M1}), 
$$\frac
{ \mbox{E} \{ Y_j^{\bar{\ba}(t)}(t) \mid \bar{\bX}(t), \bar{\mathbf{S}}(t-1), \bar{\mathbf{I}}(t-1) \}}
{ \mbox{E} \{ Y_j^{\bar{\ba}'(t)}(t) \mid \bar{\bX}(t), \bar{\mathbf{S}}(t-1), \bar{\mathbf{I}}(t-1) \}} = \exp ( \delta_1 ),$$ 
\noindent where $\bar{\bX}(t)$ is the full history of covariates and intervention through time $t$, and $\bar{\mathbf{S}}(t-1)$ and $\bar{\mathbf{I}}(t-1)$ are the full history of susceptible and infected, excluding time $t$. (Because $S_j(t-1) + I_j(t-1) + R_j(t-1) = N_j$, observing $\bar{\mathbf{S}}(t-1)$ and $\bar{\mathbf{I}}(t-1)$ implies $\bar{\mathbf{R}}(t-1)$.) Then $\exp ( \delta_1 )$ entails the risk ratio of new cases by increasing mobility level by 1 unit locally (direct effect). If $v_j(t)$ is nonzero but small in comparison to $\beta_j^{\bar{\mathbf{a}}(t)}(t)\exp\{\theta_j(t)\}$, this should hold as a good approximation.

Similarly, if we consider two regimes $\bar{\ba}(t)$ and $\tilde{\ba}'(t)$ where all components are the same except for $\bar{a}_j(t) = \bar{a}_j'(t)+1$, and $v_j(t) = 0$, then 
$$\frac
{ \mbox{E} \{ Y_j^{\bar{\ba}(t)}(t) \mid \bar{\bX}(t), \bar{\mathbf{S}}(t-1), \bar{\mathbf{I}}(t-1) \}}
{ \mbox{E} \{ Y_j^{\bar{\ba}'(t)}(t) \mid \bar{\bX}(t), \bar{\mathbf{S}}(t-1), \bar{\mathbf{I}}(t-1) \}} = \exp ( \delta_2 )$$
encodes the risk ratio of new cases by increasing mobility level by 1 unit non-locally (indirect effect). 

For the effect from neighbor $k$, consider two regimes $\bar{\ba}(t)$ and $\bar{\ba}'(t)$, where all components are the same, except for $a_k(t)$ and $a'_k(t)+1$, and $v_j(t) = 0$.  Then 
$$\frac { \mbox{E} \{ Y_j^{\bar{\ba}(t)}(t) \mid \bar{\bX}(t), \bar{\mathbf{S}}(t-1), \bar{\mathbf{I}}(t-1) \}} { \mbox{E} \{ Y_j^{\bar{\ba}'(t)}(t) \mid \bar{\bX}(t), \bar{\mathbf{S}}(t-1), \bar{\mathbf{I}}(t-1) \}} = \exp (  \delta_2 c_{jk}/m_j ),$$which encodes the indirect effect of area $k$ on area $j$.

To link the potential variables and the observed variables and identify the direct and indirect effects, we invoke the following assumptions.

\begin{assumption}[Interference] \label{a:interference}
Given the past information through $t$, the intervention $\bA(t)$ affects the potential infection rate at time $t$ and region $j$ only through $A_j(t)$ and $\tilde{A}_j(t)$.
\end{assumption}

\begin{assumption}[Causal Consistency] \label{a:consistency}
$Y_j(t) = Y_j^{\bar{\bA}(t)}(t),~ \forall t,j$.
\end{assumption}

\begin{assumption}[Ignorability of intervention process variables]  \label{a:ignorability} 
$\\ A_j(t), \tilde{A}_j(t) \ind Y_j^{\bar{\ba}(t)}(t) \mid \{ \bar{\bX}(t), \bar{\mathbf{S}}(t-1), \bar{\mathbf{I}}(t-1) \},~ \forall t,j$.
\end{assumption}

To facilitate recovery of causal estimates, we define a generalized propensity score for the direct/indirect effects
\begin{align*}
\bar{e}_{jt} \{ \kappa_1, \kappa_2; \bar{\bX}(t), \bar{\mathbf{S}}(t-1), \bar{\mathbf{I}}(t-1) \}
= f\{ A_j(t) = \kappa_1, \tilde{A}_j(t) = \kappa_2,  \mid \bar{\bX}(t)
, \bar{\mathbf{S}}(t-1), \bar{\mathbf{I}}(t-1)\}.
\end{align*}
In practice, we often assume that the direct and indirect interventions are independent, and can write this score as the two separate components
\begin{align*}
e_{jt}\{ \kappa; \bar{\bX}(t), \bar{\mathbf{S}}(t-1), \bar{\mathbf{I}}(t-1) \}
= f\{ A_j(t) = \kappa \mid \bar{\bX}(t), \bar{\mathbf{S}}(t-1), \bar{\mathbf{I}}(t-1)  \}, \\
\tilde{e}_{jt}\{ \kappa; \bar{\bX}(t), \bar{\mathbf{S}}(t-1), \bar{\mathbf{I}}(t-1) \}
= \tilde{f}\{ \tilde{A}_j(t) = \kappa \mid \bar{\bX}(t), \bar{\mathbf{S}}(t-1), \bar{\mathbf{I}}(t-1)  \},
\end{align*}
denoted by $e_{jt}\{ \kappa; \bar{\bX}(t), \bar{\mathbf{S}}(t-1), \bar{\mathbf{I}}(t-1) \}$ and $\tilde{e}_{jt}\{ \kappa; \bar{\bX}(t), \bar{\mathbf{S}}(t-1), \bar{\mathbf{I}}(t-1) \}$, respectively.

Lastly we require a positivity assumption that ensures all regions can take any plausible mobility level. This is formulated based on the propensity score:
\begin{assumption}[Positivity] \label{a:positivity}
For all $\bar{\bX}(t)$, $\bar{\mathbf{S}}(t-1)$, $\bar{\mathbf{I}}(t-1)$ with $f \{ \bar{\bX}(t), \bar{\mathbf{S}}(t-1), \bar{\mathbf{I}}(t-1) \}>0$, $\bar{e}_{jt}\{ \kappa_1, \kappa_2; \bar{\bX}(t), \bar{\mathbf{S}}(t-1), \bar{\mathbf{I}}(t-1)  \}>0, \forall t, \kappa_1, \kappa_2$ where $f(\cdot)$ denotes the generic probability function; i.e., it is a probability density function for a continuous variable and a probability mass function for a discrete variable. 
\end{assumption}

Under Assumptions \ref{a:consistency} and \ref{a:ignorability}, the induced model from (\ref{eq:M1}) for $Y_j(t)$ is $Y_j(t) \sim $ \\ $\text{Poisson} \{ \lambda_j (t) \}$. One can fit the induced model with the observed data to estimate the causal parameters. Because $\bar{\bX}$, $\bar{\mathbf{I}}$, and $\bar{\mathbf{S}}$ are high-dimensional, fitting the above model directly may become a daunting task. We consider the generalized propensity score approach for dimension reduction. Under Assumption \ref{a:ignorability}, and following \cite{rosenbaum1983central} or Imbens' generalized propensity score approach, we can show that 
\begin{align}
\label{eq:balancing}
A_j(t), \tilde{A}_j(t) \ind \{ \bar{\bX}(t), \bar{\mathbf{S}}(t-1), \bar{\mathbf{I}}(t-1) \} \mid \bar{e}_{jt} \{ \kappa_1, \kappa_2; \bar{\bX}(t), \bar{\mathbf{S}}(t-1), \bar{\mathbf{I}}(t-1) \}, \forall \kappa_1, \kappa_2. 
\tag{Balancing}    
\end{align}

In practice, following \cite{giffin2020generalized} we model the two generalized propensity scores using linear regressions of $A_j(t)$ and $\tilde{A}_j(t)$ onto covariates and past interventions and responses. Then, their distributions for a given $j$ and $t$ (i.e., their generalized propensity scores values) can be completely summarized with a single sufficient statistic: the estimated conditional mean (i.e., the fitted value from the regression). For this reason, henceforth $e_j(t)$ and $\tilde{e}_j(t)$ (without the index for $\kappa$) will refer to these sufficient conditional means, for which the result (\ref{eq:balancing}) applies. This shows that the generalized propensity score serves as a balancing score; i.e., at the same level of the propensity score, the distribution of the history of confounding variables are the same across different intervention levels.

Allowing $\bX$ and $\tilde{\bX}$ to include propensity score components, the induced model for $Y_j(t)$ given the intervention and generalized propensity score is
\begin{align*}
\mbox{E}
\{ Y_j(t) \mid &\bar{\bA}(t), \bar{\be}(t), \bar{\mathbf{Y}}(t-1) \} = 
\exp \{ A_j(t)\delta_1 +\tilde A_j(t)\delta_2 + \alpha_1+\bX_j(t)^\top \alpha_2 +  \tilde\bX_j(t)^\top \alpha_3\} \\
&\qquad\qquad\qquad\cdot \exp \{\theta_j (t)\}  + v_j(t).
\end{align*} 
The derivation is provided in the Appendix. Therefore, we can fit the model $Y_j(t)\sim$ Poisson$\{ \lambda^e_j(t) \}$, where 
\begin{align} 
\lambda_j^{e}(t) &=  \beta_j^{e}(t)  \exp \{\theta_j(t)\} + v_j(t), \\ 
\beta_j^{e}(t) &= 
\exp \{ A_j(t)\delta_1 +\tilde A_j(t)\delta_2 + \alpha_1+\bX_j(t)^\top \alpha_2 +  \tilde\bX_j(t)^\top \alpha_3\}
,  \nonumber 
\end{align} 

To model the propensity of $A_t$ given the past history, we require some structural assumption to reduce the dimension of the historical variables. We posit a model for $A_j(t)$ adjusting for its direct causes $A_j(t-1)$, $A_j(t-2)$, $X_j(t)$, $X_j(t-1)$ and $Y_j(t-1)$ to get $e_j(t)$. $\tilde{A}_j(t)$ can be modeled similarly with the neighbor-averaged variables $\tilde{A}_j(t-2)$, $\tilde{X}_j(t)$, $\tilde{X}_j(t-1)$ and $\tilde{Y}_j(t-1)$.  In practice, simply taking the average over neighbors of $e_j(t)$ gives similar estimates for $\tilde{e}_j(t)$.

\section{Simulation study}\label{s:sim}

In this section we conduct a simulation study to evaluate the statistical properties of the causal effects that stem from the approximate SIR models. Data are generated from the full SIR model in Section \ref{s:SIR} and analyzed using the approximated spatial model in Section \ref{s:approx}. The objectives are to determine when the approximations give unbiased point estimation and valid interval estimation, and to compare different approximations using these criteria.  

\subsection{Methods}\label{s:sim:methods}

We generate data on a $15\times 15$ square grid of $J=225$ regions for $T=30$ time steps.  Rook neighbors are considered adjacent.  Each region has population $N_j=100,000$ and initial states are generated as $I_j(1)=100\exp\{U_j\}$, $R_j(1)=0$ and $S_j(1)=N_j-I_j(1)$, where $(U_1,...,U_J)\sim\mbox{CAR}(1,\rho_s)$.  A confounding variable $X_j(t)$ is  generated from the STCAR$(1,\rho_s,\rho_t)$ distribution and the intervention variable is generated as $A_j(t) = \rho_xX_j(t) + \sqrt{1-\rho_x^2}E_j(t)$ where $E_j(t)$ is also generated from the STCAR$(1,\rho_s,\rho_t)$ distribution. The latent states for times $t\in\{2,...,T\}$ are generated from the full mechanistic model given by (\ref{e:SIR}) and (\ref{e:lambda}) with recovery rate $\gamma=0.1$ and infection rate $$\log\{\beta_j(t)\} = \alpha_0 + X_j(t)\alpha_1+{\tilde X}_j(t)\alpha_2 + A_j(t)\delta_1+{\tilde A}_j(t)\delta_2$$ for $\alpha_0=-3$, $\alpha_1=0.5$, $\alpha_2=0.3$, $\delta_1=0.5$ and $\delta_2=0.2$.  The data are then generated as $Y_j(t)\sim\mbox{Poisson}\{p\lambda_j(t-l)\}$ where the reporting rate is $p=0.5$ and the lag is $l=2$.

We compare six simulation scenarios defined by the spatial ($\rho_s$) and temporal ($\rho_t$) dependence of the intervention variable $A_j(t)$, the strength of confounding $(\rho_x)$ and the strength of spatial dependence ($\phi$) in the SIR model:
\begin{enumerate}
  \item Base model: $\rho_s=0.9$, $\rho_t=0.5$, $\rho_x=0.5$ and $\phi=0.4$
  \item Strong spatial dependence in $A_j(t)$: $\rho_s=0.99$, $\rho_t=0.5$, $\rho_x=0.5$ and $\phi=0.4$
  \item Strong temporal dependence in $A_j(t)$: $\rho_s=0.3$, $\rho_t=0.9$, $\rho_x=0.5$ and $\phi=0.4$  
  \item Strong spatiotemporal dependence $A_j(t)$: $\rho_s=0.9$, $\rho_t=0.9$, $\rho_x=0.5$ and $\phi=0.4$
  \item Strong confounding: $\rho_s=0.9$,
  $\rho_t=0.5$, $\rho_x=0.9$ and $\phi=0.4$
  \item Weak SIR spatial dependence: $\rho_s=0.9$,
  $\rho_t=0.5$, $\rho_x=0.5$ and $\phi=0.2$\end{enumerate}
For each scenario we generate 100 datasets.  

\hspace{-20mm}
\begin{table}[h!]
\caption{{\bf Spatiotemporal model simulation study results}. The six data-generating scenarios are given in Section \ref{s:sim:methods}.  The two recommended models, the full model (``Full'') and the model without a nugget term (``No nugget''), are given in bold; and are compared to the model without the generalized propensity score (``No PS'') and the model without spatial dependence (``Non-spatial'') in terms of bias and empirical coverage of 90\% prediction intervals, separately for the direct $(\delta_1$) and indirect $(\delta_2)$ effects. Bias is  multiplied by 100 and standard errors are given in subscripts.}\vspace{6pt} \label{t:sim}
\footnotesize
\centering
\begin{tabular}{ p{1.8in} l|cc|cc}	
& & \multicolumn{2}{c|}{Direct effect} & \multicolumn{2}{c}{Indirect effect}   \\
Scenario & Method & Bias & 90\% Cov. & Bias & 90\% Cov. \\
 \hline
\multirow{4}{*}{ \begin{tabular}{l} 1. Base Model \end{tabular} }
  & \textbf{Full} 
  & \textbf{0.25$_{0.25}$} 
  & \textbf{90$_{3}$} &  \textbf{0.36$_{0.45}$} 
  & \textbf{93$_{3}$} \\ 
  & \textbf{No nugget}
  &  \textbf{0.12$_{0.24}$} 
  & \textbf{94$_{2}$} &  \textbf{0.36$_{0.46}$} 
  & \textbf{94$_{2}$} \\ 
  & No PS 
  & 2.18$_{0.25}$ 
  & 74$_{4}$ &  2.19$_{0.53}$ 
  & 79$_{4}$ \\ 
  & Non-spatial 
  & 0.52$_{0.25}$ 
  & 96$_{2}$ & -0.41$_{0.64}$ 
  & 80$_{4}$ \\  
\hline
\multirow{4}{*}{ \begin{tabular}{l} 2. Strong spatial\\~~~~~~dependence in $A_j(t)$ \end{tabular} }
  & \textbf{Full} 
  &  \textbf{0.79$_{0.24}$} 
  & \textbf{92$_{3}$} &  \textbf{0.25$_{0.44}$} 
  & \textbf{91$_{3}$} \\ 
  & \textbf{No nugget} 
  &  \textbf{0.60$_{0.23}$} 
  & \textbf{91$_{3}$} &  \textbf{0.09$_{0.43}$} 
  & \textbf{92$_{3}$} \\ 
  & No PS 
  &  2.87$_{0.23}$ 
  & 65$_{5}$ &  1.64$_{0.54}$ 
  & 78$_{4}$ \\ 
  & Non-spatial 
  &  1.59$_{0.26}$ 
  & 99$_{1}$ & -1.02$_{0.71}$
  & 73$_{4}$ \\ 
\hline 
\multirow{4}{*}{ \begin{tabular}{l} 3. Strong temporal\\~~~~~~dependence in $A_j(t)$ \end{tabular} }
  & \textbf{Full} 
  &  \textbf{1.16$_{0.47}$} 
  & \textbf{95$_{2}$} &  \textbf{0.62$_{0.96}$} 
  & \textbf{93$_{3}$} \\ 
  & \textbf{No nugget} 
  &  \textbf{1.06$_{0.47}$} 
  & \textbf{94$_{2}$} &  \textbf{0.73$_{0.94}$} 
  & \textbf{93$_{3}$} \\ 
  & No PS 
  & 12.71$_{0.45}$  
  &  2$_{1}$ & 14.28$_{1.01}$ 
  & 20$_{4}$ \\ 
  & Non-spatial 
  &  1.78$_{0.53}$ 
  & 97$_{2}$ & -0.06$_{1.29}$ 
  & 83$_{4}$ \\ 
\hline 
\multirow{4}{*}{ \begin{tabular}{l} 4. Strong spatiotemporal\\~~~~~~dependence in $A_j(t)$ \end{tabular} }
  & \textbf{Full} 
  &  \textbf{2.15$_{0.53}$} 
  & \textbf{86$_{4}$} &  \textbf{3.63$_{1.04}$}  
  & \textbf{86$_{3}$} \\ 
  & \textbf{No nugget} 
  &  \textbf{2.18$_{0.53}$} 
  & \textbf{88$_{3}$} &  \textbf{3.43$_{1.02}$} 
  & \textbf{86$_{3}$} \\ 
  & No PS 
  & 13.59$_{0.51}$ 
  &  4$_{2}$ & 17.37$_{1.26}$ 
  & 22$_{4}$ \\ 
  & Non-spatial 
  &  4.44$_{0.62}$ 
  & 94$_{2}$ &  2.15$_{1.63}$ 
  & 72$_{5}$ \\ 
\hline 
\multirow{4}{*}{ \begin{tabular}{l} 5. Strong confounding \end{tabular} }
  & \textbf{Full} 
  & \textbf{0.97$_{0.46}$} 
  & \textbf{92$_{3}$} & \textbf{-0.86$_{1.01}$} 
  & \textbf{91$_{3}$} \\ 
  & \textbf{No nugget} 
  & \textbf{0.77$_{0.46}$} 
  & \textbf{94$_{2}$} & \textbf{-0.90$_{0.98}$} 
  & \textbf{92$_{3}$} \\ 
  & No PS 
  &  2.99$_{0.50}$ 
  & 78$_{4}$ &  1.44$_{1.09}$ 
  & 82$_{4}$ \\ 
  & Non-spatial 
  &  1.33$_{0.49}$ 
  & 95$_{2}$ & -0.31$_{1.37}$ 
  & 78$_{4}$ \\ 
\hline 
\multirow{4}{*}{ \begin{tabular}{l} 6. Weak SIR spatial\\~~~~~~dependence \end{tabular} }
  & \textbf{Full} 
  & \textbf{0.28$_{0.25}$} 
  & \textbf{92$_{3}$} &  \textbf{0.67$_{0.48}$} 
  & \textbf{96$_{2}$} \\ 
  & \textbf{No nugget} 
  & \textbf{0.22$_{0.25}$} 
  & \textbf{93$_{3}$} & \textbf{0.55$_{0.48}$} 
  & \textbf{96$_{2}$} \\ 
  & No PS 
  &  2.53$_{0.26}$ 
  & 74$_{4}$ &  1.91$_{0.56}$ 
  & 82$_{4}$ \\ 
  & Non-spatial 
  &  0.64$_{0.27}$ 
  & 97$_{2}$ & -0.30$_{0.67}$ 
  & 80$_{4}$ \\ 
\hline   
\end{tabular}
\end{table}

The propensity score $e_j(t)$ is the fitted value from a least squares regression of $A_j(t)$ onto $A_j(t-1)$, $A_j(t-2)$, $X_j(t)$, $X_j(t-1)$ and $Y_j(t-1)$. We then model $\log\{\beta_j(t)\}$ as a linear combination of $A_j(t)$, ${\tilde A}_j(t)$, $X_j(t)$, ${\tilde X}_j(t)$, $e_j(t)$, ${\tilde e}_j(t)$, $e_j(t)^2$ , ${\tilde e}_j(t)^2$ and $e_j(t){\tilde e}_j(t)$.   For each simulated dataset we fit the full model in (\ref{e:finalmodel}) with the propensity scores included as covariates (``Full''), and compare this model to the full model but with $v_j(t)=0$ (``No nugget''), the full model but without the propensity scores $e_j(t)=\tilde{e}_j(t)=0$ (``No PS'') and the full model but without spatial dependence $\rho_s=0$ (``Non-spatial'').  For the non-spatial model we also set $v_j(t)=0$ because the MCMC algorithm did not converge with these terms included.  The models are fit using responses from time points $t\in\{5,...,n_t\}$ to accommodate lagged predictors.  We fit the model using MCMC with 10,000 iterations and the first 2,000 samples are discarded as burn-in to approximate the posterior median and 95\% interval of the direct $(\delta_1)$ and indirect $(\delta_2)$ effects.

\subsection{Results}\label{s:sim:results}

The results are given in Table \ref{t:sim}. Broadly speaking the Full model and the No Nugget model perform well. Giving similar precision and coverage results under all scenarios, they generally outperform the other models in terms of coverage and bias. Moreover, their coverage rates remain reasonably close to their 90\% target, although they appear to have a small, positive bias. The Non-spatial model performs worse than the Full and No Nugget models in both precision and coverage, but it remains competitive. The No PS model performs the worst, with extremely poor performance for the strong temporal/spatiotemporal dependence scenarios. The four models seem to have the most difficulty estimating effects when there is strong temporal/spatiotemporal dependence in $A_j(t)$ or strong confounding. All models appear to estimate the direct effects better than the indirect effects.

\section{Estimating the causal effect of community mobility reduction on Coronavirus spread}\label{s:app}

We implement similar models that were used in the simulation study on the real data: A Full model, a No Nugget model, a No PS model, and a Non-spatial model. The models are fit using responses from time points $t\in\{8,...,31\}$ (April 24, 2020 through October 8, 2020) to accommodate lagged predictors. The direct propensity score is estimated with least squares regressions of $A_j(t)$  onto the previous intervention $A_j(t-1)$, the local covariates $X_j(t)$ and $X_j(t-1)$, the number of weeks since the first case in county $j$ (enters both linearly and quadratically), time $t$ (enters both linearly and quadratically), a time and intervention interaction $t \cdot A_j(t-1)$, the baseline level of mobility $A_j(1)$, and $\log\{Y_j(t-1)\} - \log(N_j)$. The propensity score $e_j(t)$ is set to the fitted values of the resulting model. The propensity of the spillover intervention is estimated as the local average of its neighboring direct scores: $\tilde{e}_j(t) = \sum_{k \sim j}e_k(t)/m_j$. The direct and indirect propensity scores are added to $\bX$ and $\tilde{\bX}$, respectively, as in (\ref{e:betaFormInFullSpec}), along with $l$-lagged interventions and covariates. Each model is implemented with lags of $l = 0,\ldots, 7$.  The No Nugget, No PS, and Non-spatial models were run for 100,000 MCMC iterations, with a burn-in tuning period of 20,000 iterations. The Full model required more iterations to converge, so these values were increased to 200,000 and 40,000, respectively.

\subsection{Effect estimation}\label{s:app:results}

\begin{figure} 
\normalsize
  \centering 
  \includegraphics[width=.9\linewidth]{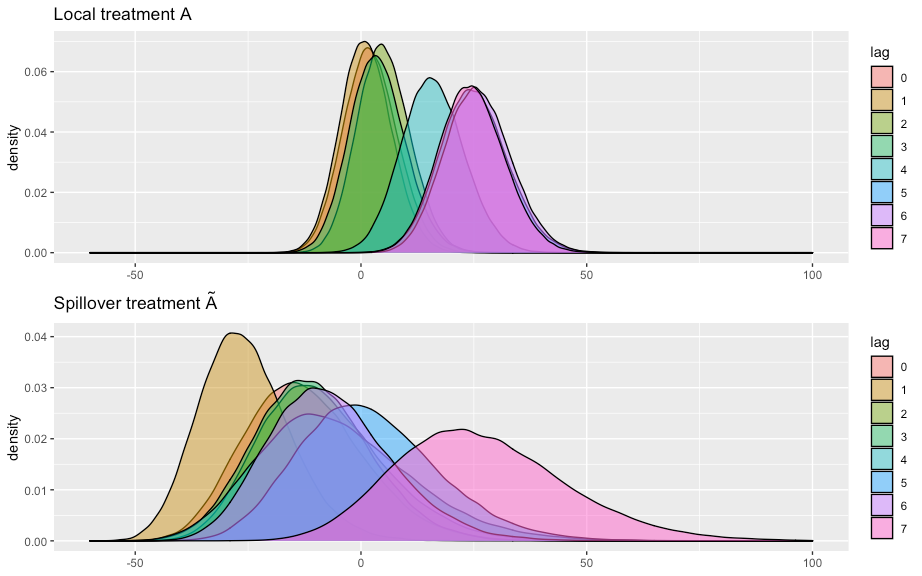}
  \caption{\textbf{Posterior distributions for direct indirect intervention effects for Full model across lags 0-7 weeks.} Estimates correspond to the expected percentage increase in cases from an increase in mobility 50\% above baseline: $100\cdot\{\exp(50\cdot\delta_i)-1\}$, $i=1,2$. The direct effect estimates for lags 3-5 are 3.5, 15.7, and 24.8, with the latter two significant at the $\alpha = 0.05$ level.  The indirect estimates for lags 3-5 are -9.6, -7.6, and 0.3, though none are significant.}
  \label{f:densities}
\end{figure}

Table \ref{t:estsOverLags} gives the estimated posteriors for the direct and spillover effects for the four different models across lags $l = 0,\ldots,7$, and Figure \ref{f:densities} shows these results graphically for the Full model. 
Rather than present the estimated values for $\delta_1$ and $\delta_2$, the estimates shown are $100\cdot\{\exp(50\cdot\delta_i)-1\}$, $i=1,2$, and correspond to the expected percentage increase in cases from an increase in mobility 50\% above baseline.   
Because it is not clear that the observed data can identify the correct lag, we provide model results for all lags, and speculate that lags of 3-5 weeks are most appropriate to see effects of changes in mobility. The Centers for Disease Control and Prevention guidelines suggest that COVID symptoms appear 2-14 days after virus exposure \citep{CDC}. Given this, 3-5 weeks would allow the virus to spread through roughly several generations of people -- which would be enough to see changes in county-level case counts.

\begin{table}[h!]
\caption{{\bf Posterior estimates for direct and indirect treatment effects for lags 0-7 weeks for all models.} 
  Estimates correspond to the expected percentage increase in cases from an increase in mobility 50\% above baseline ($100 \cdot (\exp(50\cdot\delta)-1)$).  
  95\% confidence intervals are shown in parentheses, and estimates significant at the 95\% level are denoted with $*$.}\vspace{6pt} \label{t:estsOverLags}
\centering
\begin{tabular}{lc|cc|cc}
Model & Lag (weeks) & \multicolumn{2}{c}{Direct effect} & 
\multicolumn{2}{c}{Indirect effect} \\
\hline 
\multirow{8}{*}{ \begin{tabular}{l} Full \end{tabular} }
&0&2.0& (-9.0, 14.1) \phantom{*}&-12.9& (-35.1, 17.0)\phantom{*}\\  
&1&1.0& (-9.5, 12.6) \phantom{*}&-28.0& (-45.7, -4.6)*\\  
&2&4.9& (-5.8, 16.8) \phantom{*}&-9.4& (-31.8, 22.1)\phantom{*}\\    
&3&3.5& (-7.8, 16.4) \phantom{*}&-9.6& (-32.5, 20.0)\phantom{*}\\    
&4&15.7& (2.8, 29.9) *&-7.6& (-33.1, 28.3)\phantom{*}\\    
&5&24.8& (11.0, 40.2) *&0.3& (-25.3, 34.6)\phantom{*}\\    
&6&25.7& (12.3, 40.7) *&-9.9& (-33.2, 21.0)\phantom{*}\\    
&7&24.3& (11.0, 39.1) *&24.2& (-7.5, 66.5)\phantom{*}\\    
\hline
\multirow{8}{*}{ \begin{tabular}{l} No Nugget\end{tabular} }
&0&-0.8& (-11.6, 11.7) \phantom{*}&-21.7& (-41.1, 4.1)\phantom{*}\\  
&1&-1.5& (-11.8, 9.8) \phantom{*}&-28.5& (-46.4, -4.6)*\\  
&2&3.6& (-7.2, 15.6) \phantom{*}&-15.4& (-35.7, 12.1)\phantom{*}\\    
&3&0.3& (-10.4, 12.3) \phantom{*}&-19.3& (-38.5, 6.2)\phantom{*}\\    
&4&12.3& (-0.3, 28.0) \phantom{*}&-13.7& (-35.5, 15.4)\phantom{*}\\    
&5&22.7& (9.1, 37.8) *&3.1& (-22.8, 37.7)\phantom{*}\\    
&6&22.2& (8.7, 37.2) *&-18.0& (-38.8, 11.4)\phantom{*}\\    
&7&22.8& (9.2, 37.7) *&2.3& (-22.9, 35.7)\phantom{*}\\    
\hline
\multirow{8}{*}{ \begin{tabular}{l} No PS\end{tabular} }
&0&-2.7& (-8.0, 2.8) \phantom{*}&2.9& (-9.4, 17.1)\phantom{*}\\  
&1&2.1& (-3.1, 7.7) \phantom{*}&8.7& (-5.0, 24.1)\phantom{*}\\  
&2&6.6& (0.9, 12.9) *&20.1& (5.1, 39.2)*\\    
&3&11.9& (5.6, 18.7) *&32.5& (12.3, 57.3)*\\    
&4&15.6& (9.2, 22.3) *&32.7& (15.1, 59.5)*\\    
&5&15.3& (9.6, 21.6) *&27.6& (10.4, 48.4)*\\    
&6&12.4& (6.7, 18.4) *&20.3& (5.6, 37.4)*\\    
&7&10.8& (5.2, 16.7) *&21.0& (6.3, 37.2)*\\    
\hline
\multirow{8}{*}{ \begin{tabular}{l} Non-spatial\end{tabular} }
&0&-2.6& (-17.0, 13.8) \phantom{*}&7.7& (-12.5, 32.0)\phantom{*}\\  
&1&-5.9& (-19.8, 10.1) \phantom{*}&-46.4& (-57.1, -34.5)*\\  
&2&-0.9& (-14.9, 15.5) \phantom{*}&-25.7& (-39.3, -9.4)*\\    
&3&-4.3& (-17.7, 11.5) \phantom{*}&-37.2& (-48.3, -23.5)*\\    
&4&9.8& (-6.1, 28.8) \phantom{*}&-21.6& (-35.2, -4.2)*\\    
&5&25.1& (7.3, 45.5) *&35.8& (13.3, 62.4)*\\    
&6&19.3& (3.0, 38.0) *&-10.5& (-24.3, 6.2)\phantom{*}\\    
&7&24.2& (6.8, 44.5) *&6.1& (-11.1, 26.5)\phantom{*}\\    
\hline 
\end{tabular}
\end{table} 

The key trend evident in these results is that the direct effect appears to be positive and significantly different from zero for higher lags. The Full model gives positive significant estimates on $\delta_1$ for lags 4-7 weeks; the No Nugget for lags 5-7 weeks; the No PS model for lags 2-7 weeks; and the Non-spatial model for lags 5-7. The size of this effect differs between models/lags, but the Full model for lags 5-7 predicts that a mobility level 50 percent above baseline will increase Coronavirus cases by between 11 and 41 percent. This trend is demonstrated in the grouping of posteriors above zero in the top sub-figure of Figure \ref{f:densities}. This provides evidence that a decrease in local mobility should decrease local Coronavirus cases roughly 4-7 weeks after the change in mobility. 

The indirect effect estimates are more variable with wider credible intervals, as illustrated in Figure \ref{f:densities}. For the Full model, a lag of 7 gives a positive but non-significant estimate for the indirect effect, although the indirect effect posteriors for other lags seem centered at zero, with the exception of lag 1 which is significant and negative.  For lag 1 the No Nugget model also shows this counter-intuitive result. The No PS model shows a number of positive and significant indirect effects, and the Non-spatial model paradoxically shows several significant indirect effects -- some positive and some negative. It is not entirely clear why there is more variability in the indirect effects or why we see these counter-intuitive results.  Some of these could simply be spurious significance resulting from estimating many lags. Another possibility is that our measure of connectivity does not precisely measure the levels of contact between different counties. This could explain the wide intervals, as the connectivity is over-estimated between some counties and under-estimated between others.  In any event, the indirect effect story appears less clear cut than that of the direct effect. 

Table \ref{t:regLag4} gives parameter estimates from the No Nugget/Lag 5 model. In addition to the intervention effects and accounting for the propensity scores, the covariate coefficients tell an interesting story about their association with Coronavirus case counts. For example, expected Coronavirus case counts are decreasing in median age of population, population density, county population, and relative humidity. Alternatively, Coronavirus case counts tend to increase with the poverty rate, PM$_{2.5}$, number of hospital beds, number of ICU beds, and percentage healthcare-employed residents.

\begin{table}[h]
\caption{{\bf Posterior estimates for coefficients in No Nugget, lag 5 model.}  95\% confidence intervals are shown in parentheses, and estimates significant at the 95\% level are 
  denoted with $*$. Coefficients are not transformed as in Table \ref{t:estsOverLags}.}\vspace{6pt} \label{t:regLag4}
\centering
\begin{tabular}{r|c}
Variable & Estimate (95\% interval)\\
\hline 
Intercept & -7.43 (-7.68, -7.18)     * \\
$A_j(t)$ & 0.004 (0.002, 0.006)     * \\
$\tilde{A}_j(t)$ & 0.0003 (-0.0055, 0.006) \\
$e_j(t)$ &  -0.0011 (0.0007, 0.0054)     * \\
$\tilde{e}_j(t)$ &  0.006 (0.001, 0.013)     * \\
Mean temperature & -0.008 (-0.016, 0.000)     * \\
Relative Humidity & -0.009 (-0.011, -0.006)     * \\
Dewpoint & 0.045 (0.033, 0.057)     * \\
PM$_{2.5}$ & 0.082 (0.066, 0.097)     * \\
Poverty rate & 1.07 (0.88, 1.26)     * \\
Population density & -0.035 (-0.043, -0.026)     * \\
Median household income (thousands) & -0.002 (-0.003, -0.001) * \\
Population (millions) & -0.048 (-0.066, -0.030) * \\
\# hospital beds & 5.95 (3.64, 8.29)     * \\
\# ICU beds & 24.01 (6.99, 40.94)     * \\
Median age & -0.021 (-0.023, -0.018)     * \\
Proportion of foreign-born residents & 4.16 (3.97, 4.35)     * \\
Proportion of health-employed & 0.82 (0.59, 1.04)     * \\
$\rho_s$ & 0.9949 (0.9945, 0.9950)     * \\
$\rho_t$ & 0.452 (0.441, 0.463)     * \\
\hline 
\end{tabular}
\end{table}

\section{Discussion}\label{s:discussion}

This analysis uses a spatiotemporal model to provide insights into how mobility affects the spread of Coronavirus. The analysis uses a propensity score framework to obtain causal estimates of this effect, while allowing for potential interference between nearby counties. We show via simulation study that the causal estimates from our computationally-efficient approximate model have good statistical properties for data generated from a mechanistic process. In particular, the data suggest that decreases in mobility appear to cause a statistically significant decrease on the number of local Coronavirus cases after roughly 5 weeks. This is significant, as such a long lag might be overlooked in a naive analysis of the relationship between mobility and Coronavirus. 

There are several noteworthy limitations of this analysis. The intervention data are somewhat crude measures of mobility. Taken from a small minority of smartphone users, they provide a limited view into the distribution of mobility among residents for a given county and week. Moreover, the five relevant measures of mobility provided by Google (which we average over) may have different intervention effects.  We also do not have data on other interventions to reduce the spread of Coronavirus, such as wearing masks, state-mandated closures of different types of businesses, social distancing, washing hands, etc. These are all plausibly correlated with mobility, so it is possible that the intervention effect of mobility incorporates the causal effect of these measures too. 

The measure of connectivity that we use -- that of adjacency between counties -- is also somewhat crude.  For example, while the first viral hotspot in the United States is thought to have been Seattle, it quickly travelled to New York by air. A more comprehensive measure of connectivity and travel between regions may be helpful to incorporate to the model.

The response data also has limitations. Because the availability of Coronavirus tests in the United States has varied over time and space, the number of new positive tests results in a particular county/week might reflect both the prevalence of Coronavirus, but also the availability of tests. One solution to this would be to use COVID-19-related hospitalizations or deaths as a response. However, these are necessarily less common -- providing fewer response data -- and they would add to the already substantial lag between the moment of infection and positive test.

Finally, as we make the ``no unmeasured confounders'' assumption, it is possible that there are key confounders that have not been included. Because of this, we include a number of possible confounders, but more can always be done in this regard.  

As the pandemic continues, many of these data issues are being addressed: surveys are now being done to ascertain levels of mask-usage and other social distancing practices. Publicly available data on the availability of Coronavirus tests by location are emerging.  Further, as the number of cases continues to grow, using hospitalizations or deaths as a response becomes more feasible.

\section*{Acknowledgements}
The authors thank Sukanya Bhattacharyya, Can Cui, Matt Miller, Parker Trostle, Laura Wendelberger, Stephen Xu and Zun Yin (North Carolina State University), Yawen Guan (University of Nebraska) and Pulong Ma (Duke University) for help compiling the data. This work was partially supported by NIH grant R01ES031651-01.
\bibliographystyle{apalike}
\bibliography{refs}

\section*{Disclaimer: The views expressed in this manuscript are those of the individual authors and do not necessarily reflect the views and policies of the U.S. Environmental Protection Agency. Mention of trade names or commercial products does not constitute endorsement or recommendation for use.}

\newpage
\section*{Appendix: Derivation of conditional expectation of $Y$}\label{s:A3b}

\begin{align*}
\mbox{E}
\{ Y_j(t) \mid &\bar{\bA}(t), \bar{\be}(t), \bar{\mathbf{Y}}(t-1), \bar{\bX}(t) \} 
\\
&= \mbox{E}
\{ Y_j(t) \mid \bar{\bA}(t), \bar{\be}(t), \bar{\mathbf{S}}(t-1), \bar{\mathbf{I}}(t-1), \bar{\bX}(t) \} 
\\
&= \mbox{E}\left[ \mbox{E}\{ Y_j(t) \mid \bar{\bA}(t),\bar{\bX}(t), \bar{\mathbf{S}}(t-1), \bar{\mathbf{I}}(t-1)\}  \mid \bar{\bA}(t),\bar{\be}(t), \bar{\mathbf{S}}(t-1), \bar{\mathbf{I}}(t-1), \bar{\bX}(t)\right] 
\\
&= \mbox{E}\left[ \mbox{E}\{ Y_j^{\bar{\bA}(t)}(t) \mid \bar{\bA}(t),\bar{\bX}(t), \bar{\mathbf{S}}(t-1), \bar{\mathbf{I}}(t-1)\}  \mid \bar{\bA}(t),\bar{\be}(t), \bar{\mathbf{S}}(t-1), \bar{\mathbf{I}}(t-1), \bar{\bX}(t)\right] 
\\
&= \mbox{E} \{ \lambda_j^{\bar{\bA}(t)}(t) \mid \bar{\bA}(t),\bar{\be}(t), \bar{\mathbf{S}}(t-1), \bar{\mathbf{I}}(t-1) , \bar{\bX}(t)\} 
\\
&= \exp \{ A_j(t)\delta_1 +\tilde A_j(t)\delta_2\} \cdot
\mbox{E} \left[\exp \{ \alpha_1+\bX_j(t)^\top \alpha_2 +  \tilde\bX_j(t)^\top \alpha_3\} \bigg| \bar{\be}(t), \bar{\mathbf{S}}(t-1), \bar{\mathbf{I}}(t-1), \bar{\bX}(t) \right]
\\
&\qquad\qquad \cdot \exp \{\theta_j (t)\}  + v_j(t)   \\
&= 
\exp \{ A_j(t)\delta_1 +\tilde A_j(t)\delta_2 + \alpha_1+\bX_j(t)^\top \alpha_2 +  \tilde\bX_j(t)^\top \alpha_3\} \cdot \exp \{\theta_j (t)\}  + v_j(t)   ,
\end{align*}
Note that the conditional expectation on the last line is a function of $\bar{\mathbf{e}}(t)$. The second equality follows from the causal consistency assumption, the third from the ignorability assumption on the intervention process, the fourth from the form of model (\ref{eq:M1}), and the fifth from the balancing property (\ref{eq:balancing}) of the generalized propensity score. The final equality assumes that the generalized propensity scores components have been included in $\bX$ and $\tilde{\bX}$.
\end{document}